# Three-Pass Protocol Implementation in Vigenere Cipher Classic Cryptography Algorithm with Keystream Generator Modification

Amin Subandi[*,1], Rini Meiyanti[1], Cut Lika Mestika Sandy[1], Rahmat Widia Sembiring[2]

[1]Faculty of Computer and Information Technology, " Universitas Sumatera Utara", Medan, 20155, Indonesia,

[2]Politeknik Negeri Medan, Medan, 20155, Indonesia,

Emails: aminsubandi@yahoo.com, inimeiyanti522@gmail.com, likasandy3@gmail.com, rahmatws@polmed.ac.id

A B S T R A C T

*Vigenere Cipher is one of the classic cryptographic algorithms and included into symmetric key cryptography algorithm, where to encryption and decryption process use the same key. Vigenere Cipher has the disadvantage that if key length is not equal to the length of the plaintext, then the key will be repeated until equal to the plaintext length, it course allows cryptanalysts to make the process of cryptanalysis. And weaknesses of the symmetric key cryptographic algorithm is the safety of key distribution factor, if the key is known by others, then the function of cryptography itself become useless. Based on two such weaknesses, in this study, we modify the key on Vigenere Cipher, so when the key length smaller than the length of plaintext entered, the key will be generated by a process, so the next key character will be different from the previous key character. In This study also applied the technique of Three-pass protocol, a technique which message sender does not need to send the key, because each using its own key for the message encryption and decryption process, so the security of a message would be more difficult to solved.*

## 1. Introduction

A security issue is one of the most important aspects of an information system. A message that contains important information can be misused by irresponsible people, therefore a message contains important information, in order to secure an important message it is necessary a technique to secure it, cryptography is the science and art to maintain the security of a message [11].

The development of cryptography itself has already begun a long time ago, there are two types of cryptography, classical and modern cryptography, classical cryptography works based on character mode, and modern cryptography works based on bit mode. And if viewed from the key, cryptographic algorithm using symmetric and asymmetric key do not revise any of the current designations.

[*]Corresponding Author: Amin Subandi, Faculty of Computer and Information Technology, Universitas Sumatera Utara, Medan, 20155, Indonesia
Email: aminsubandi@yahoo.com

www.astesj.com

Vigenere Cipher is a classic cryptographic algorithm, classical cryptography is generally included into the symmetric key algorithm, where to do the encryption and decryption process use the same key. In this case, key security and key distribution become the main factor, when the key and the ciphertext is known, then, of course, plaintext will be known also. This is one of the drawbacks of symmetric key algorithms.

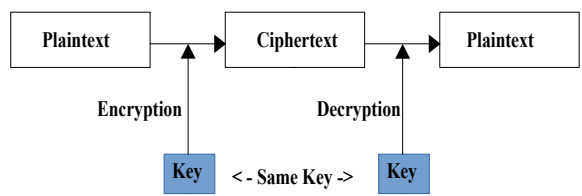

**Figure 1:** Symmetric key algorithm scheme

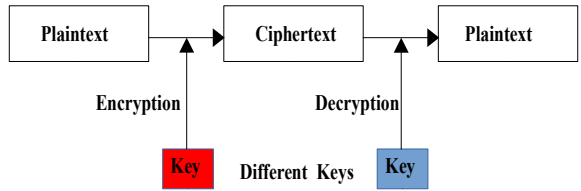

**Figure 2:** Asymmetric key algorithm scheme

Vigenere Cipher itself also has drawbacks which have been solved by a kasiski method, The drawback is if the key length is not the equal to the length of plaintext, then the key will be repeated continuously until the same as the plaintext length, it can cause the occurrence of what is called the histogram, which is the same ciphertext or repetitive content, kasiski method collect all the repeated histogram to calculated the distance between the histogram to find the length of the initial key.

With the existence of these flaws, we interested to try and examine how to minimize weaknesses. And we try to modify Vigenere Cipher algorithm with the Key Stream Generator method, which if key length is not the same length of the plaintext, then the next key will be generated by a process, this will cause key to getting the same length as the plaintext length becomes unrepetitive and it is expected to thus it would be more difficult to solve by kasiski method.

Furthermore, this paper also will apply what the so-called Three-Pass protocol, a method for securing messages without having to distribute keys, because neither the sender nor the recipient can use owned key each to encryption and decryption process. It is expected also that by applying the layered method (Vigenere Cipher key modification and use Three-Pass protocol methods), can increase the security level of classical cryptography algorithms Vigenere Cipher.

## 2. Theories

### 2.1. Three-Pass Protocol

The Three-Pass protocol is a framework that allows a party may send a message encrypted securely to the other party without having provided the key [2], this is possible because between the sender and receiver using a key belonging to the respective to perform the encryption and decryption process. Called the Three-pass protocol as do three exchange time before a message decrypted into meaningful messages. Three-Pass protocol invented by Adi Shamer about 1980 [2]. in applying Three-Pass protocol does not always have to use a cryptographic algorithm, because basically, this technique has its own function, namely to use the function exclusive-OR (XOR) [1], but in practice, to improve the reliability of this technique combined with cryptographic algorithms.

The implementation of this technique is still less attention [1], In the previous study, This technique can be a solution to the classic problem of the symmetric key algorithm which should send a key to the recipient. A message sender only needs to send a message to the recipient, and the important one (the key) does not need to distribute.

Here is a schematic representation Three-Pass protocol:

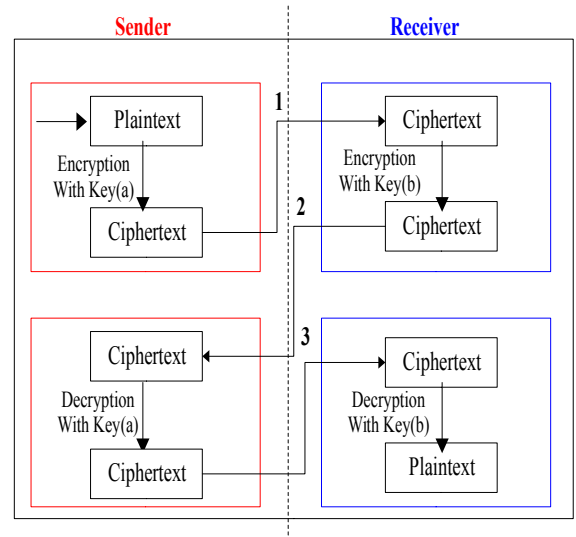

**Figure 3:** Three-Pass Protocol Process Scheme

### 2.2. Vigenere Cipher

Vigenere Cipher is one of the classic cryptographic algorithms that included into the category of polyalphabetic substitution [3] and a symmetric key cryptographic algorithm, whereby for encryption and decryption process used same keys. In the process of encryption and decryption, Vigenere Cipher using a table called tabula recta [11], it is a 26 x 26 matrix containing alphabet letters. This Algorithm was discovered by Blaise de Vigenere of France in the century to sixteen, in 1586 and this algorithm cannot be solved until 1917 [3] [11] Friedman and Kasiski solved it [3].

**Figure 4**: Vigenere Cipher table

Mathematically, the process of encryption and decryption Vigenere Cipher can be seen in the following equation:

$$C_i = E ( P_i + K_i ) \bmod 26 \quad (1)$$

$$P_i = D ( C_i - K_i ) \bmod 26 \quad (2)$$

Which *C* is the ciphertext, *P* is the plaintext, *K* is the Key, *E* is Encryption than *D* is Decryption. For the encryption process, every plaintext alphabet combined with the alphabet keys, the alphabet which intersected (by table) between the plaintext and ciphertext is the alphabet ciphertext. If the key alphabet is smaller than the plaintext, then the key will be repeated until equal to the length of the plaintext.

For example, supposing that the plaintext "THIS IS MY PAPER" with the keyword "UP", then, based on the table square vigenere, illustrations encryption can be seen as follows:

```
Plaintext   : THIS IS MY PAPER
Keyword     : UPUP UP UP UPUPU
Ciphertext  : NWCH CH GN JPJTL
```

For the decryption process, with vigenere cipher table and used the same key, then the resulting plaintext from the ciphertext comparison with the keys, the alphabet corresponding to the key and the ciphertext, then the alphabet is the plaintexts.

```
Ciphertext  : NWCH CH GN JPJTL
Keyword     : UPUP UP UP UPUPU
Plaintext   : THIS IS MY PAPER
```

The drawback of Algorithm Vigenere Cipher is if the key length is smaller than the plaintext length, then the key will be repeated, because it most likely will produce the same ciphertext as long as the same plaintext, in the example above, the character "IS" in the encryption into ciphertext the same as "CH", this can be exploited by cryptanalysts to break off the ciphertext. Kasiski method is a method of cryptanalysis that broke Vigenere Cipher algorithms, methods kasiski collect the same characters of ciphertext to calculate the distance to ultimately find a number of the key length. After a long key is found, the next step is determining what the keywords to use exhaustive key search [11].

*2.3. Keystream Generator*

KeyStream Generator is a process for generating the key by using a function, so that be randomly generated the key. With the random key, it will further increase the reliability of a cryptographic algorithm because it would be difficult to solve.

The function used may be any function by use of bait as an input that is not the same for each process, so it will produce different output depending on that input.

Process keystream generator receives the initial fill of *U* (as a user key), and then processed to generate keys $K_i$, the next insert is from the previous process, and so on until the process reached $K_n$.

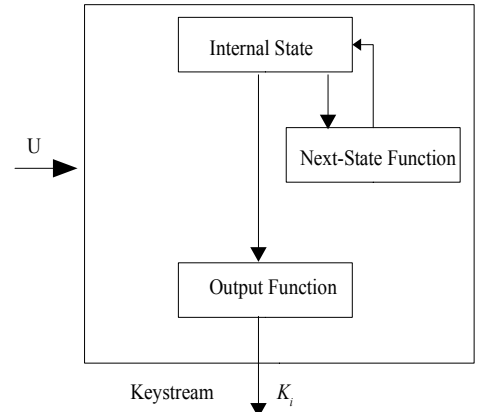

**Figure 5:** Key Stream Generator Process

## 3. Methodology

In this paper, it will be implemented how to implement cryptographic algorithms classic Vigenere Cipher by the key modification and how it is applied in the Three-pass protocol that the sender and the receiver do not have to exchange keys.

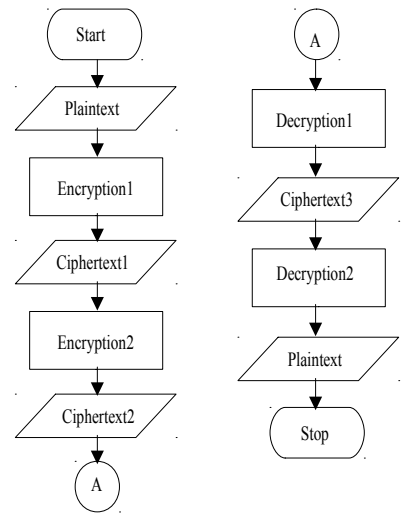

**Figure 6:** Flowchart of Three-Pass Protocol scheme

The key modification of Vigenere Cipher in this study, we use the key stream generator, the key will be generated when the key length is not equal to the length of the plaintext, when the key length equal to the length of the plaintext, the key generating process is not necessary.

In generating the keys, we use the following equation:

$$K_i = (K_{i-1} + n) \bmod 26 \quad (3)$$

where $K_i$ is a character key that will be generated, $K_{i-1}$ is the index of the previous key character and *n* is the length of the keys to the $K_{i-1}$.

Alphabet in the alphabet *A to Z* can be described as an index of a row of numbers from *0 to 25* with *0 = A* and *25 = Z*, an illustration of the process of generating the key is illustrated as follows: suppose the length of plaintext is 10, and the length of the is key 6, suppose the key text "MYCODE", when the standard key Vigenere Cipher, key will be "MYCODEMYCO", then by using the above functions, the whole key generate to "MYCODEKRZI", where *"K"* is generated from the summation of index *"E"= 4* plus the length of the key of "M" to "E"= 6 equals 10 then in modulo with 26 generating a character to *10 ="K"*, and so on.

For encryption and decryption is then performed as usual. Flowchart for encryption and decryption process vigenere cipher can be seen as follows:

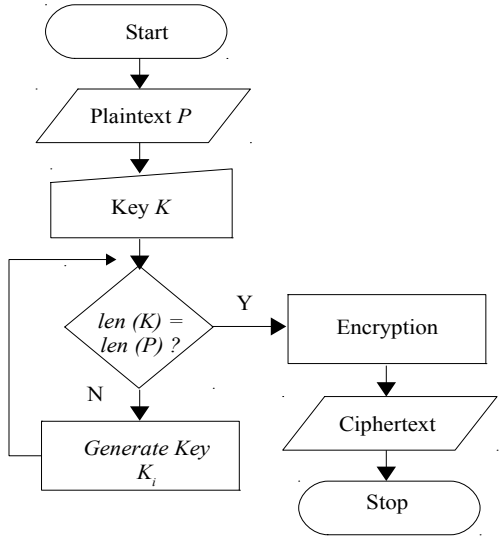

**Figure 7:** Vigenere Cipher Encryption Process

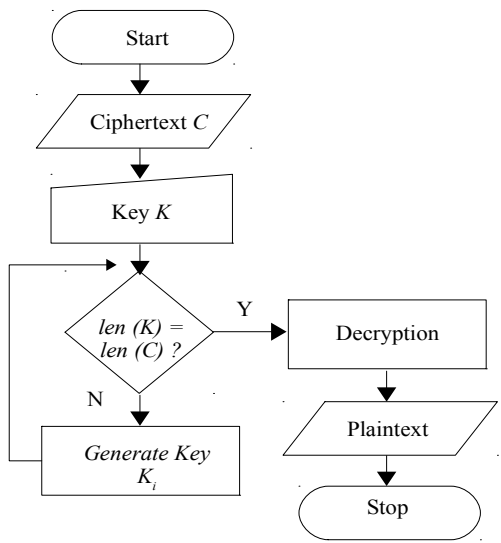

**Figure 8:** Vigenere Cipher Decryption processed

In this paper, to prove each process, we use python, we made it to both encryption and decryption process in plaintext manual input or plaintext taken from a file with .txt extension.

## 4. Testings And Implementations

As a test, we will do the encryption and decryption of a message with a cryptographic algorithm classic Vigenere Cipher that reads "THE FAMILY AND THE FAV" uses two keys are assumed as a key owned by the sender "KEY" and keys owned by the recipient "BUNG".

First, the sender will encrypt the plaintext with his own key and produce the first ciphertext, then the first ciphertext sent to the recipient, the recipient encrypts back the first ciphertext with his own key and produce the second ciphertext, and then the second ciphertext is sent back by the recipient to the sender, then sender decrypt the second ciphertext and generates the third ciphertext, the third ciphertext is sent back to the recipient to decrypt into plaintext.

For the illustrations can be seen as follows:

**First Encryption by Sender:**
Plaintext            : THE FAMILY AND THE FAV
Sender Key           : KEY BFKQXF OYJ VIW LBS
First Ciphertext     : DLC GFWYID OLM OPA QBN

**Second Encryption by Recipient:**
First Ciphertext     : DLC GFWYID OLM OPA QBN
Recipient Key        : BUN GKPVCK TDO ANB QGX
Second Ciphertext    : EFP MPLTKN HOA OCB GHK

**First Decryption by Sender:**
Second Ciphertext    : EFP MPLTKN HOA OCB GHK
Sender Key           : KEY BFKQXF OYJ VIW LBS
Third Ciphertext     : UBR LKBDNI TQR TUF VGS

**Second Decryption by Recipient:**
Third Ciphertext     : UBR LKBDNI TQR TUF VGS
Recipient Key        : BUN GKPVCK TDO ANB QGX
Plaintext            : THE FAMILY AND THE FAV

In the above process, we can see that there have been three exchanges between the sender and the recipient, the first exchange is the sender sends the first ciphertext, the second exchange is the recipient sends back the second ciphertext to the sender, the last exchange is the sender sends the third ciphertext, and finally the recipient decrypts the message to be plaintext.

For comparison, let's see how if the message decrypted by standard Vigenere cipher algorithm method. Where the sender sends the key to the recipient for decrypt the ciphertext simultaneously through different paths and assuming that the key delivered safely.

**Encryption by Sender:**
Plaintext           : THE FAMILY AND THE FAV
Key                 : KEY KEYKEY KEY KEY KEY
Ciphertext          : DLC PEKSPW KRB DLC PET

**Decryption by Recipient:**
Ciphertext          : DLC PEKSPW KRB DLC PET
Key                 : KEY KEYKEY KEY KEY KEY
Plaintext           : THE FAMILY AND THE FAV

For encryption process using the key that has been modified, it appears that the ciphertext generated more random, the word "THE" produce different cipher word, it will be different if we decrypt without key modification, the word "THE" will be decrypted the same word "DLC", then "FA" equal to "PE", this will be loopholes for cryptanalyst to make the cryptanalysis process using kasiski methods. By modifying the key, kasiski method will be more difficult to do, moreover, the key also has to distributed, this, of course, requires us to ensure that the distribution of the key should be completely safe. But when we applying the method of Three-Pass protocol, the key does not need to be distributed.

## 5. Conclusions

From the research, it seemed that by modifying keys, classical algorithm Vigenere Cipher actually has better reliability compared with standard Vigenere Cipher, this is due to the modification of the keys that are generated from a process that is done, so that when the key length is not equal to the length of the plaintext, then the key will not be repeated, but will be generated by a function, this has resulted in a more random keys rather than having to repeat the key as in the standard Vigenere Cipher algorithm.

And Vigenere Cipher can also be applied to the method of Three-Pass protocol, so although Vigenere Cipher included into the algorithm symmetric key, the sender of the message does not have to send the key used to encrypt the message, because each can use its own key both to encrypt and decrypt, of course, this is very useful when the key distribution security more vulnerable to tapping, and moreover, if the distribution of key is secure, the message sent does not have to be encrypted, isn't it ?.

In this study, we simply apply safeguards messages using the standard alphabet consists of 26 characters, for further research, may be applied to a more complex character.